\documentclass[extra,mreferee,a4wide]{article}
\usepackage{graphicx}
\usepackage{color}

\newcommand{\be}{\begin{equation}}
\newcommand{\ee}{\end{equation}}
\newcommand{\bea}{\begin{eqnarray}}
\newcommand{\eea}{\end{eqnarray}}

\title {Earthquake statistics inferred from plastic events in soft-glassy materials.}

\author{Roberto Benzi$^1$, Federico Toschi$^2$ and Jeannot Trampert$^3$\\
$^{1}$Department of Physics, University of ''Tor Vergata``, \\Via della Ricerca Scientifica 1, 00133 Rome, Italy. \\
$^{2}$Department of Physics, Eindhoven University of Technology, \\PO Box 513, 5600  MB, Eindhoven, The Netherlands. \\
$^{3}$Department of Earth Sciences, Utrecht University, \\ PO Box 80115, NL-3508 TC, Utrecht,The Netherlands.
}

\begin{document}

\maketitle
\begin{abstract}
We propose a new approach for generating synthetic earthquake catalogues based on the physics of soft glasses. The continuum approach produces yield-stress materials based on Lattice-Boltzmann simulations. We show that, if the material is stimulated below yield stress, plastic events occur, which have strong similarities with seismic events. Based on a suitable definition of displacement in the continuum, we show that the plastic events obey a Gutenberg-Richter law with exponents similar to those for real earthquakes. We further find that average acceleration, energy release, stress drop and recurrence times scale with the same exponent. The approach is fully self-consistent and all quantities can be calculated at all scales without the need of ad hoc friction or statistical laws. We therefore suggest that our approach may lead to new insight into understanding of the physics connecting the micro and macro scale of earthquakes.
\end{abstract}

\section{Introduction}

It is commonly accepted that earthquakes are the result of some mechanical failure of earth materials. However, many details of the underlying physics, especially at the microscopic scale, are currently not understood. At the macroscopic scale, many aspects of earthquakes show a complex behaviour that can be addressed with tools of statistical physics. Indeed, earthquakes obey empirical power laws possessing a certain scale invariance. The best known laws are the Gutenberg-Richter (GR) law \cite{gb:1954}, which relates the frequency of earthquakes to their magnitude or seismic moment (energy) and Omori's law \cite{omori:1894, utsu_etal:1995}, which describes how the frequency of aftershocks decays with time. There are other empirical laws \cite{turcotte_etal:2007}, but they are much less well documented. The reason for this power law behaviour of earthquakes is still under debate. 
Power laws and scale invariance have been central in statistical physics in the context of self-organised criticality. If the earth is in a permanent critical state due to inherent dynamics, self-organised criticality could explain the power law behaviour of earthquakes. Certain aspects of earthquakes, however, are better represented by characteristic earthquakes giving rise to characteristic energy and time scales. There is also some evidence that 'mode-switching' between dynamical regimes occurs. The latter behaviour can be understood in terms of generalised phase changes between discrete and continuum states of material. For an overview of these discussions see \cite{turcotte_etal:2007, ben_zion:2008}.

Deciding between scale invariance or characteristic scales is a difficult problem because the dynamics involved are not directly observable. It is mostly approached by generating synthetic earthquake catalogues and comparing those to the observed GR law and/or Omori's law from recorded earthquake catalogues. These synthetic catalogues are generated by purely statistical models, purely mechanical models, or a mixture of both. An overview of the commonly employed models can be found in \cite{turcotte_etal:2007} and \cite{verejones:2011}. The ETAS (epidemic-type aftershock sequence) model, for instance,  is a popular statistical model, whereas the block-slider model is a popular mechanical one where everything is determined by the prescribed friction law \cite{kawamura_etal:2012}. The block-slider model is most often implemented numerically \cite{rundle_etal:2003}, whereas laboratory based studies are less common \cite{rubinstein_etal:2011}. None of these synthetic earthquake models provides a fundamental insight into the physics of earthquakes at the microscopic scale, which is ultimately responsible for the empirical laws seen for real earthquakes. The sequences are generated by ad-hoc prescribed statistical or constitutive laws.

Since the pioneering study of \cite{mogi:1962}, acoustic emissions of fracture experiments in the laboratory were monitored and their power law behaviour studied. An overview of earlier studies can be found in \cite{rundle_etal:2003}. Laboratory based fracture experiments are performed under controlled dynamic fracture regimes and the rupture can be observed from inception to arrest. \cite{yoshimitsu_etal:2014} showed that there is a natural continuation in the scaling between seismic moment and corner frequency from kilometer-size natural earthquakes and millimeter-scale micro-ruptures in rocks.

We propose a new approach for generating synthetic earthquakes catalogues which  is based on the physics of complex soft-glassy materials. This continuum approach, based on the Lattice-Boltzmann method, provides the way to simulate  yield-stress materials (i.e. viscous fluid-like behaviour above a critical yield stress and plastic solid-like below the critical yield stress) \cite{benzi_etal:2010}. 
Below yield stress, plastic events can be identified radiating elastic perturbations through the model \cite{benzi_etal:2014} similar to earthquakes. In the present paper, we show that these plastic events follow the GR law with b-values comparable to those for observed earthquake sequences. Our approach is not based on any phenomenological laws but on the exact momentum equations of a mixture of immiscible  fluids. All physical properties can thus be computed at any scale \cite{benzi_etal:2014, benzi_etal:2015}. It is hoped that a comparison with real earthquake data can yield insight into the physics responsible for the observed empirical laws.  We remark that, recently (see for instance \cite{Liu:2015tk}, \cite{Salerno:2013dz} and \cite{Lin:2014hz} ) a similar approach has been proposed using molecular dynamics simulations for glass forming systems. 

\section{Generating plastic events and their analysis}

We consider a model of soft glass recently introduced into the literature \cite{benzi_etal:2009} and based  on a lattice kinetic description. 
The basic idea of the model
is to consider two non-ideal fluids with particular frustration effects at the interface in order to stabilize phase separation against coarsening. In particular,
we start from a mesoscopic lattice Boltzmann model for non-ideal binary fluids, which combines a small positive surface tension, promoting highly complex interfaces, with a positive disjoining pressure, inhibiting interface coalescence. The mesoscopic kinetic model considers two fluids $A$ and $B$, each described by a {\it discrete} kinetic distribution function $f_{\zeta i}({\bf r},{\bf c}_i;t)$, measuring the probability of finding a particle of fluid $\zeta =A,B$ at position ${\bf r}$ and time $t$, with a discrete velocity ${\bf c}_i$, where the index $i$ runs over the nearest and next-to-nearest neighbours of ${\bf r}$ in a regular two-dimensional  lattice.  The meso-scale particle represents all molecules contained in a unit cell of the lattice. The distribution functions evolve with time under the effect of free-streaming and local two-body collisions, described, for both fluids ($\zeta=A,B$), by a relaxation towards a local equilibrium ($f_{\zeta i}^{(eq)}$) with a characteristic time-scale $\tau_{LB}$:
\begin{equation}
\label{LB}
f_{\zeta i}({\bf r}+{\bf c}_i,{\bf c}_i;t+1) -f_{\zeta i}({\bf r},{\bf c}_i;t)  = -\frac{1}{\tau_{LB}} \left(f_{\zeta i}-f_{\zeta i}^{(eq)} \right)({\bf r},{\bf c}_i;t)+F_{\zeta i}({\bf r},{\bf c}_i;t).
\end{equation}
The equilibrium distribution is given by
\begin{equation}
f_{\zeta i}^{(eq)}=w_i\rho_{\zeta} \left[1+\frac{ {\bf u}  {\bf c}_i} {c_s^2} + \frac{ {\bf u}{\bf u}:({\bf c}_i{\bf c}_i-c_s^2)} {2 c_s^4} \right]
\end{equation}
with $w_i$ a set of weights known a priori \cite{ss:2011}. Coarse grained hydrodynamical densities for both species are defined  as $\rho_{\zeta }=\sum_i f_{\zeta   i}$ and the global momentum for the whole binary mixture as ${\bf j}=\rho {\bf u}=\sum_{\zeta , i} f_{\zeta i} {\bf c}_i$, with $\rho=\sum_{\zeta} \rho_{\zeta}$. The term $F_{\zeta i}({\bf r},{\bf c}_i;t)$ is the $i$-th projection of the total internal force which includes a variety of inter-particle forces.  A delicate issue concerns the choice of the forcing term which is done as follows:
First,  we consider a repulsive ($r$) force with strength parameter ${\cal G}_{AB}$ between the two
fluids
\begin{equation}
\label{Phase}
{\bf F}^{(r)}_\zeta ({\bf r})=-{\cal G}_{AB} \rho_{\zeta }({\bf r}) \sum_{i, \zeta ' \neq \zeta } w_i \rho_{\zeta '}({\bf r}+{\bf c}_i){\bf c}_i
\end{equation}
which is responsible for the phase separation.  Both fluids are also subject to competing interactions whose role it is to provide a mechanism for {\it frustration} ($F$) of the  phase separation. In particular, we introduce two forces, namely a short-range (nearest neighbour, NN) self-attraction, controlled by strength parameters ${\cal G}_{\zeta \zeta,1} <0$ and ``long-range'' (next to nearest neighbour, NNN) self-repulsion, governed by strength parameters ${\cal G}_{\zeta \zeta,2} >0$
\begin{equation}\label{NNandNNN}
{\bf F}^{(F)}_\zeta ({\bf r})=-{\cal G}_{\zeta \zeta ,1} \psi_{\zeta }({\bf r}) \sum_{i \in NN} w_i \psi_{\zeta }({\bf r}+{\bf c}_i){\bf c}_i -{\cal G}_{\zeta \zeta ,2} \psi_{\zeta }({\bf r}) \sum_{i \in NNN} w_i \psi_{\zeta }({\bf r}+{\bf c}_i){\bf c}_i
\end{equation}
with $\psi_{\zeta }({\bf r})=\psi[\rho_{\zeta }({\bf r})]=1-e^{-\rho_{\zeta }({\bf r})}$ a suitable pseudo-potential function \cite{shan_chen:1993}. Despite their inherent microscopic simplicity, the above dynamic rules are able to promote a host of non-trivial collective effects, for a detailed discussion see \cite{benzi_etal:2009}.   By proper tuning of the phase separating interactions (\ref{Phase}) and the competing interactions (\ref{NNandNNN}), the model simultaneously achieves a small positive surface tension $\Gamma$ and a positive disjoining pressure $\Pi_d$. In short, particles sitting at the interface are subject to three different forces: an attractive one from the $NN$ particles of the same species  and {\it two} repulsive forces from the $NNN$ particles of the same species and the $NN$ particles of the other species. The two opposite repulsion forces introduce a frustration effect at
the interface which is able to dramatically slow down the coarsening effect in the system. This allows the simulations of droplets of one dispersed phase into the other,  which are stabilized against coalescence. Once the droplets are stabilized, different packing fractions and poly-dispersity of the dispersed phase can be achieved. In the numerical simulations presented in this paper, the packing fraction of the dispersed phase in the continuum phase is kept approximately equal to $90 \%$. The model provides two basic advantages whose combination is not common. On the one hand, it provides a realistic structure for the emulsion droplets, like for instance the Surface Evolver method; at the same time, due to its built-in properties, the model gives directly access to equilibrium and out-of-equilibrium shear-stresses, including both elastic and viscous contributions.

\begin{figure}
\noindent\includegraphics[width=25pc]{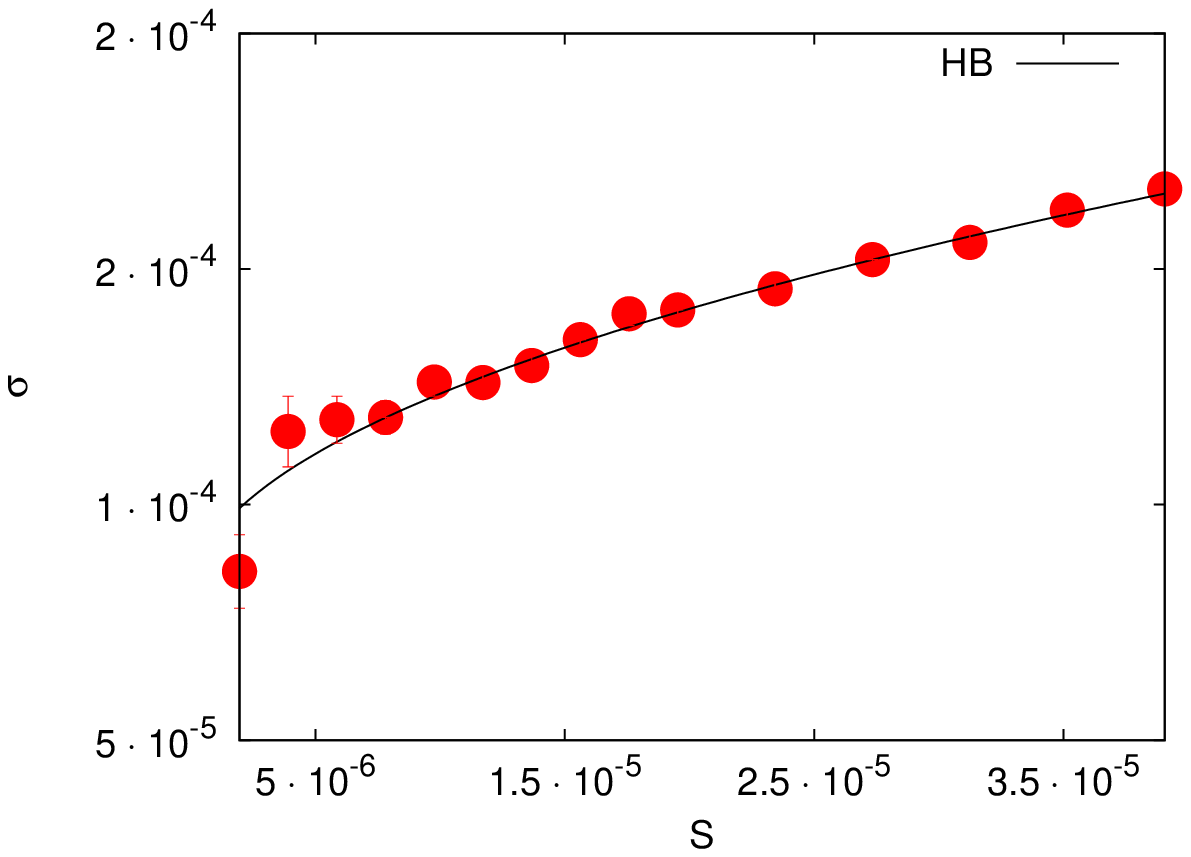}
  \caption{Shear-stress $\sigma$ as a function of the imposed strain-rate $S$ for a Couette geometry, obtained by using the Lattice-Boltzmann model described in the text. The best fitting Herschel-Bulkley law is also shown.}
\label{fig1}
\end{figure}

\begin{figure}
\noindent\includegraphics[width=35pc]{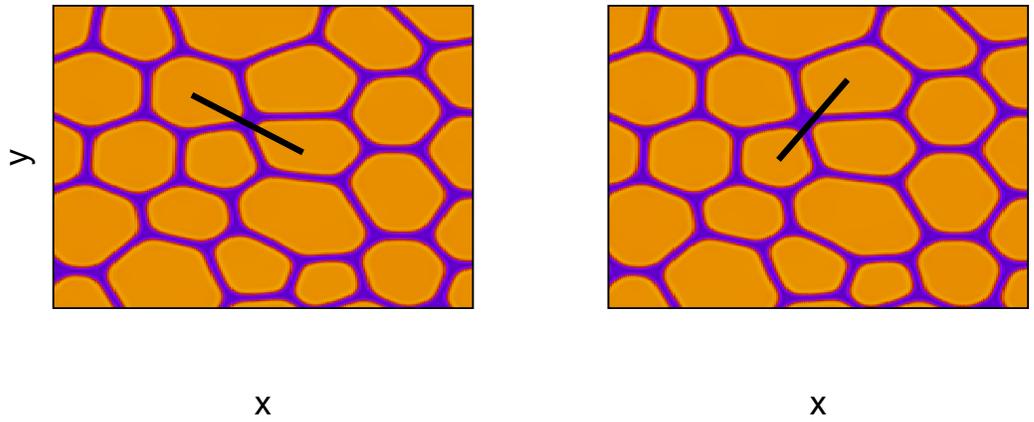}
 \caption{A typical plastic events observed in the numerical simulation of our Lattice-Boltzmann model. The figure shows a local enlargement of the density $\rho_A$ at two different times. The black lines indicate two bubbles which are in contact: from one time to the next the topological configuration of the bubbles changes.}
  \label{fig2}
\end{figure}

Upon choosing a small volume ratio between the two fluids (say the volume of fluid $A$ divided by that of fluid $B$ is small), the model exhibits a typical configuration depicted in  figure (\ref{fig2}) that closely resembles that observed in real emulsions and foams. The model shows remarkable agreement with existing experimental data, namely we observe a finite yield stress $\sigma_y$ above which the shear-stress $\sigma$ depends on the strain-rate $S$
following a Herschel-Bulkley law \cite{benzi_etal:2014, sbragaglia2015}. In figure (\ref{fig1}), we show the shear-stress $\sigma$ as a function of external strain-rate $S$ in a standard Couette geometry, where the external strain-rate is due to  the imposed boundary conditions. 

For a small external strain-rate $S$, the system does not flow. The external forcing provides the energy for isolated plastic rearrangements which usually take the form of a so called T1 events \cite{benzi_etal:2014}. In figure  (\ref{fig2}) we give an example of such plastic events. They can be isolated or multiple events can occur simultaneously. In all cases, just a few bubbles change the topological network of the system. Soon after a plastic event, elastic waves travel through the domain and can trigger other plastic events. The overall dynamics is strongly intermittent in space and time, a feature often observed in laboratory visualization of real emulsions. A standard Voronoi tessellation is able to capture the topological change in the network and the location of plastic events.  Remarkably, our model is perhaps the only one which enables the simulation of an emulsion-like system with realistic interface dynamics and with no {\it a priori} constrains on bubble sizes and shapes. 

\begin{figure}
\noindent\includegraphics[width=25pc]{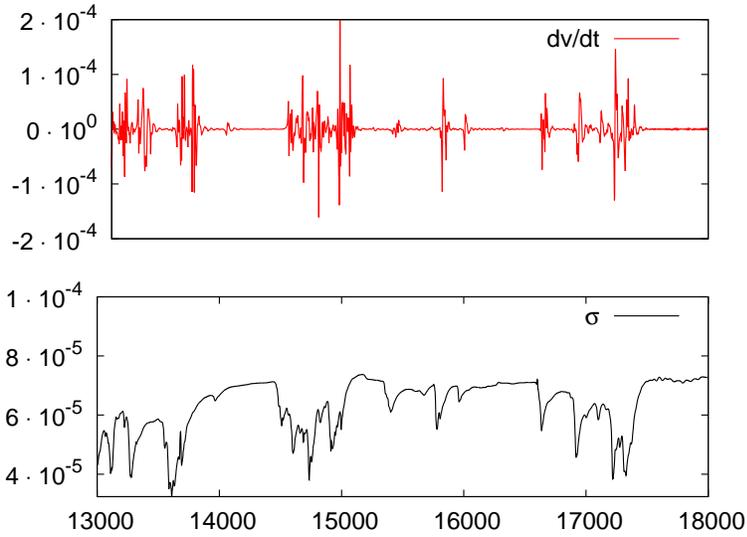}
  \caption{Behavior of $dv/dt$, upper panel, and shear-stress $\sigma$ during a relatively short time period in the numerical simulation. The system is driven at the boundaries by a very small  external strain-rate $S$ in a Couette geometry.  The velocity $v$ has been obtained as the $x$ average in the system at the center of the channel. The shear stress $\sigma$ is averaged over space in both $x$ and $y$ directions.}
  \label{fig3}
\end{figure}

For a small external strain-rate, the shear-stress is much smaller than the yield stress $\sigma_y$ and the system exhibits stick-slip behaviour. Such an intermittent stop-and-go mechanism has often been considered to be the basic mechanism underlying the statistical properties  of earthquake dynamics. Evidence for the occurrence of plastic events has been given in \cite{benzi_etal:2014}. In figure (\ref{fig3}) we show the behaviour of $\sigma(t)$  and the corresponding value of $dv/dt$ for a relatively short time window in a simulation using a Couette geometry.  $\sigma(t)$ is the space averaged stress and $v(t)$ is the velocity of the system averaged in space in the $x$ direction and computed at the centre of the channel.  In this example, we took a symmetric forcing on the two boundaries where the velocity difference $\Delta U = S L$ is fixed, $S$ being the apparent external strain-rate and $L$ the size of the system.  For this figure, we considered a system of $512^2$ grids points corresponding to about $130$ bubbles, where $S= 2.7 \times 10^{-6}$ and integrated the system for $3 \times 10^{7}$ time steps. For more information on the system equations, please see \cite{benzi_etal:2014}. In this particular case the yield stress is $\sigma_y \sim 10^{-4}$.  There are two remarkable features in figure (\ref{fig3}): first of all the stress $\sigma(t)$ intermittently shows strong drops followed by slow increases; secondly the acceleration $dv/dt$ sporadically shows with large fluctuations around a mean value of zero, reminiscent of earthquake recordings. Both effects are related to the above mentioned stick-slip mechanism. In particular, the large fluctuations in $dv/dt$ correspond to plastic events in the system, which  can be far or close from the central line where we measured $v$.  

While figure (\ref{fig3}) looks encouraging and we are tempted to associate events corresponding to strong fluctuations in $dv/dt$ to ''quakes'', the quantity $dv/dt$ is an average quantity and not suitable for a systematic investigation of the statistical properties of earthquake-like events.  In earthquake seismology the statistical properties of quakes are investigated by looking at the frequency of earthquakes of a given magnitude, which are know to follow the GR law.  To define  magnitude and seismic moment, we need to define a suitable measure of displacement. 

Let us recall that the simulation uses kinetic equations in the continuum limit. Thus we have no particle we can follow in the system. However, because the interface between the two fluids is stable, we can measure displacements by computing changes in the position of the interface. The simplest way to do this is to take our system $L \times L$ and divided it into smaller squares of size $L/n \times L/n$. We chose $n$ such that $L/n$ corresponds to $32$ grid points which is the average size of a single bubble. Furthermore we checked that our results, as discussed below, are independent of the exact choice of $n$.  Next, for each square $L/n \times L/n$ we considered two consecutive times, say $t$ and $t+\tau$ and computed the density change $\delta \rho(x,y,t,\tau) \equiv \rho(x,y,t+\tau)-\rho(x,y,t)$. Finally, we took the average of $\delta \rho(x,y,t,\tau)^2$ in square $i$, where $i$ is a label for the $n^2$ squares of sides $L/n$
and denoted it by $\delta \rho^2_i(t,\tau)$.  The reason to choose $\delta \rho^2_i(t,\tau)$ is that
for small enough  $\tau$, it easy to show that $\delta \rho^2_i(t, \tau) \sim \rho^2 {\cal A}_i(t,\tau)$ where ${\cal A}_i$ is the fraction of $n^2$ points which have been changed due to interface displacement.  
The value of ${\cal A}_i(t,\tau)$ is also proportional to $1-O_i(t,\tau)$, where $O_i(t,\tau)$ is the so-called overlap between two consecutive configurations \cite{benzi_etal:2014}. 

\begin{figure}
\noindent\includegraphics[width=25pc]{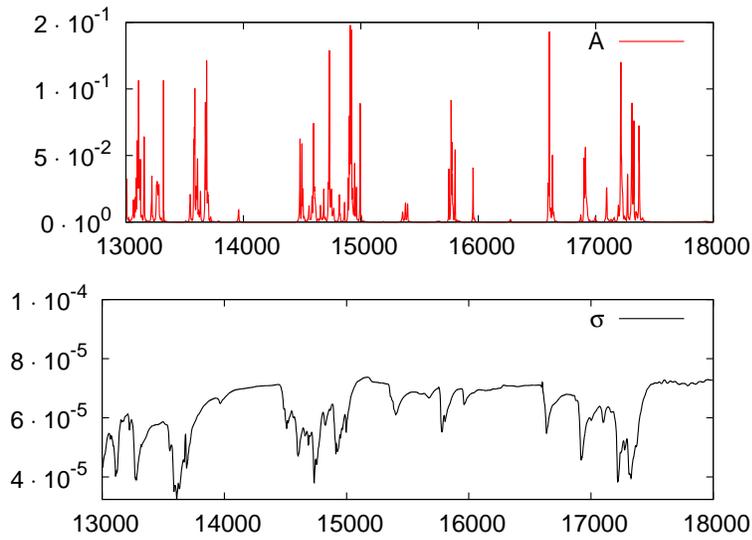}
  \caption{Same as in figure (\ref{fig3}) with the upper panel representing the quantity $A(t)$ which provides a measure of the largest displacement observed in the system.}
  \label{fig4}
\end{figure}

We now need to connect our previously defined quantities to the standard definition of earthquake magnitude or moment. The seismic moment $M_0$ is defined as $M_0 \sim D S_a$, where $D$  is the average slip of the earthquake and $S_a$ its source area. For each small square, the area $S_a$ is simply given by $(L/n)^2$ and the displacement is proportional to $ \sqrt{{\cal A}_i(t,\tau)}$. However, ${\cal A}_i (t,\tau)$ shows strong fluctuations both in space (i.e. from square  to square $i$) and in time (only at times where a plastic event occurs are one or more values of ${\cal A}_i$ relatively large). Therefore it seems reasonable to consider $ D^2 \sim A(t, \tau) = sup_i [ { \cal A}_i (t,\tau)]  $ as being representative of the squared-displacement.  Such a choice is further motivated by the fact that plastic events are local in space and are responsible for the largest value of ${\cal A}_i$ in $i$, and  we are interested to study the statistical properties of the extreme events in the displacement, which corresponds to $A(t,\tau)$. Similarly in seismology, the maximum displacement at a given frequency  is used to define a magnitude. In figure (\ref{fig4}) we show the behaviour of $A(t, \tau)$ as a function of time for the same time snapshot as already discussed in figure (\ref{fig3}).  We observe a strong correlation in the sharp increase of $A(t, \tau )$ and a drop in stress $\sigma$.  In the following we take $\tau$ to be a relatively small fraction ($0.2$) of the characteristic time scale $t_p$ for plastic events. In fact, plastic events occur over a small but non-zero value of time called $t_p$ \cite{benzi_etal:2014}. In our simulations $t_p \sim 5000$ time steps and we chose $\tau=1000$ time steps. Hereafter, we will neglect $\tau$ in the definition of $A(t)$.  The above discussion tells us therefore that we can consider  $ M_0 \sim A(t)^{1/2} $. Within the  same approximation, we can further estimate the energy release as $\Delta \sigma D S_a \sim D^2 \sqrt{S_a} \sim A(t)  $, where $\Delta \sigma $ is the stress drop during an event,  which is then proportional to the local strain $D/\sqrt{S_a}$.  

\section{Results}

Above we argued that $A(t)$ is a good candidate to investigate the statistical properties of our system.  In that case, the GR law is equivalent to a scaling behaviour of the probability density distribution of $A$ of the form:
\begin{equation}
\label{GRA}
P(A) \sim A^{-\gamma}
\end{equation}
To assess the validity of eq. (\ref{GRA}), we performed two different series of numerical simulations with resolution $512^2$ and $1024^2$ respectively. By increasing the resolution we increase the size of the system, i.e. the number of bubbles.  Numerical simulations were performed for several millions of time steps, long enough to assure the statistically invariance of  the probability density function. For each resolution we chose two different values of the external forcing with $\sigma < \sigma_y$. 

In figure (\ref{fig5}) we show a log-log plot of $P(A)$ for two different values of the strain rate $S$ for a resolution $512^2$. For both values of the strain rate a clear scaling of $P(A)$ is observed with exponents $\gamma$ in the range $[1.1,1.5]$. In order to assess the robustness of our results, we also computed the probability distribution of $dv/dt$ and of the energy release in the system, namely $E_r \sim\sigma d\sigma/dt$. 
We assume that the elastic energy in the system is proportional to $\sigma^2$ and we take compute the probability distribution of $E_r$ for $E_r<0$. 
In figure (\ref{fig6}) we show the probability distribution of $A$ for the largest strain rate of figure (\ref{fig5}) together with the probability distribution of $|dv/dt|$ and $E_r$. All quantities show almost the same scaling properties, although the range where the scaling law is observed is quantity dependent. Remarkably, the same scaling law seems to be observed for the time $\tau_E$ between two consecutive events. In particular, we defined an event when $A(t)$ is greater than a given threshold $A_{th}$ which, for figure (\ref{fig6}), is chosen to be $A_{th} = 10^{-3}$. In the insert of figure (\ref{fig6}) we show the probability distribution of $\tau_E$ where the black line corresponds to the scaling law observed in figure (\ref{fig5}). Finally in figure (\ref{fig7}), we show the probability distribution of $A(t)$ for the numerical results using a  resolution of $1024^2$. 

The results shown in figures (\ref{fig5}), (\ref{fig6}) and (\ref{fig7}) are independent of our choice of the effective displacement $D=A^{1/2}$ . We have checked that the same scaling is observed, for instance, if we
consider the quantity ${\tilde A} \equiv N_{eff}^{-1} \Sigma_i {\cal A} _i(t)$ where $N_{eff}$ is the number of  small boxes where the displacement is concentrated. More specifically, upon defining $p_i \equiv
{\cal A}_i(t) / \Sigma_i {\cal A}_i (t)$, we compute the "entropy" $H \equiv - \Sigma_i p_i log p_i$ and derived $N_{eff} \equiv \exp(H)$, which  by definition is the number of boxes where the displacement is concentrated.  It turns out that ${\tilde A} \sim A $ for all times and with very high accuracy. This supports  our choice to consider  $A$ as  an unbiased estimate of the area subject to an effective displacement.

\section{Concluding remarks}

\begin{figure}
\noindent\includegraphics[width=25pc]{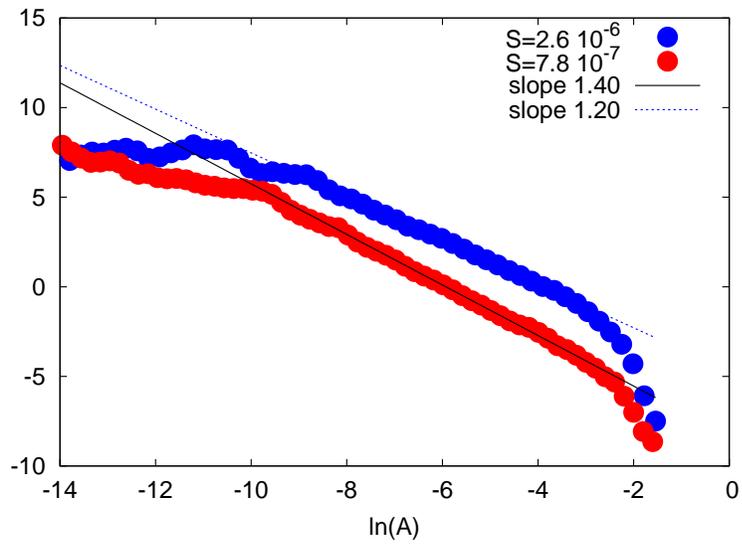}
  \caption{Log-log plot of the probability distribution $P(A)$ of $A(t)$ obtained by numerical simulations for two different values of the strain-rate. $A^{1/2}$ is proportional to the seismic moment $M_0$.}
  \label{fig5}
\end{figure}

\begin{figure}
\noindent\includegraphics[width=25pc]{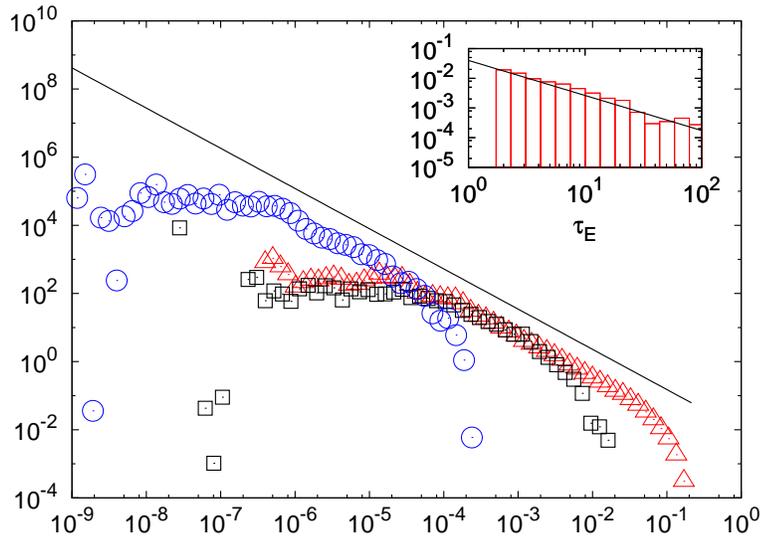}
  \caption{Probability density distribution for different quantities: red triangles correspond to the probability density of $A$ and is the same as the one shown in figure (\ref{fig5}) for the larger strain rate; blue circles correspond to the probability density function of $|dv/dt|$, where $v$ is the average velocity in the $x$ direction at the center of the channel; the black squares corresponds to the probability distribution of the energy release $E_r$. In the insert we show the probability density function of $\tau_E$ defined as the time between two consecutive events. We define an event by the condition $A_{th}>10^{-3}$. The black line in the insert has the same slope as the black line in the main part of the figure.}
  \label{fig6}
\end{figure}

\begin{figure}
  \noindent\includegraphics[width=25pc]{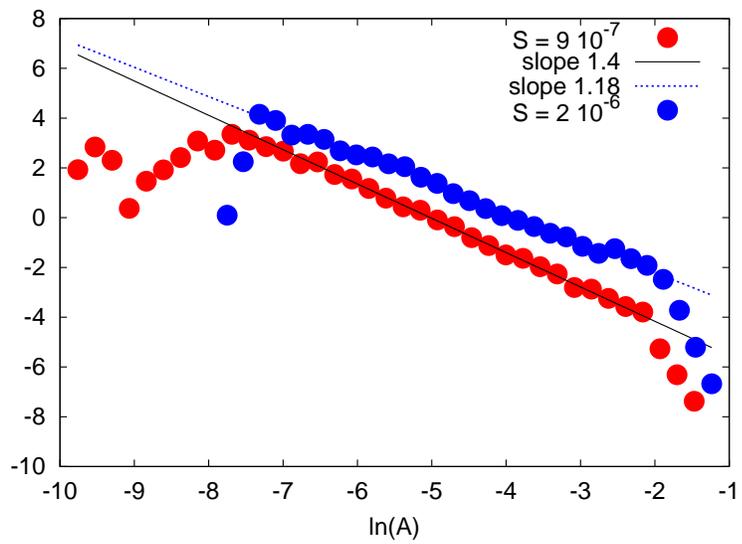}
  \caption{Same as in figure (\ref{fig5}) for a simulation at a resolution of $1024^2$. Notice that by increasing the resolution we are increasing the system size, i.e. the number of bubbles.}
\label{fig7}
\end{figure}

The information obtained from figures (\ref{fig5}),(\ref{fig6}) and (\ref{fig7}) is quite clear and striking: the system shows well defined scaling laws  for various quantities and independent of the volume considered (i.e. independent on the number of bubbles). In particular, our results strongly support  the following conclusions:

\begin{itemize}
\item { A clear scaling behaviour of $P(A)$ versus $A$ is observed };
\item { The scaling exponent lies in the range $[1.2:1.4]$ and is smaller for larger values of the external strain rate $S$};
\item { Last but not least, the scaling behaviour does not depend on the numerical resolution and it is the same for the acceleration $|dv/dt |$, the energy release $E_r$ and hence the stress drop, and the recurrence time $\tau_E$}.
\end{itemize}

To compare the values of our scaling exponents with the ones observed in the GR law for earthquakes, we have to remember that the moment $M_0 \sim D S$. In our case $D \sim A^{1/2}$ since $A$  is a measure of an area, namely  the number of unit squares subject to displacement.  If $P(A) \sim A^{-\gamma}$ then the quantity $M_0 \sim D \sim A^{1/2} $ shows a probability distribution $P(M_0) \sim M_0^{1-2\gamma}$. Upon defining $B_{GR}$ as $P(M_0) \sim M_0^{-(1+B_{GR})} $,  the results shown above give an estimate  of  $B_{GR}$ in the range $[0.4:0.8]$ not far from the scaling  reported for real earthquakes.

It is interesting to observe that the scaling exponent decreases for increasing shear rate $S$. Eventually, for very large $S$, we expect the stress to overcome the yield stress and, at that point, the system starts to flow. Only in the region $\sigma<\sigma_y$ does the system show stick-slip behaviour and GR statistics. It is worth stressing that the statistical properties of $P(A)$ are resolution independent. Last but not least,  we would like to remark that our model was not designed to mimic any realistic seismic environment and/or a particular form of friction law, yet shows all characteristics of natural earthquakes.

\bibliography{./gr_jt}
\bibliographystyle{plain}

\end{document}